\documentclass[amssymb,twocolumn,superscriptaddress]{revtex4}

\usepackage[T1]{fontenc}
\usepackage[utf8]{inputenc}
\usepackage{graphicx}
\usepackage{amsmath}
\usepackage{amssymb}
\usepackage{amsbsy}
\usepackage{times}
\usepackage[hidelinks]{hyperref}
\usepackage{graphicx}
\usepackage{xcolor}
\usepackage{siunitx}
\DeclareSIUnit\sq{\ensuremath{\Box}}  
\usepackage{textcomp}
\usepackage{natbib}
\usepackage{epstopdf}
\usepackage{nicefrac}
\usepackage{ulem} 

\renewcommand{\emph}[1]{\textit{\textbf{#1}}}

\begin{document}
\title{Strongly non-linear superconducting silicon resonators}
\author{P. Bonnet}
\author{F. Chiodi}
\email[Corresponding author: ]{francesca.chiodi@c2n.upsaclay.fr}
\affiliation{Université Paris-Saclay, CNRS, Centre de Nanosciences et de Nanotechnologies, 91120, Palaiseau, France}
\author{D. Flanigan}
\affiliation{Quantronics group, Universit{\'e} Paris-Saclay, CEA, CNRS, SPEC, 91191 Gif-sur-Yvette, France}
\author{R. Delagrange}
\author{N. Brochu}
\author{D. Débarre}
\affiliation{Université Paris-Saclay, CNRS, Centre de Nanosciences et de Nanotechnologies, 91120, Palaiseau, France}
\author{H. le Sueur}
\email[Corresponding author: ]{helene.le-sueur@cea.fr}
\affiliation{Quantronics group, Universit{\'e} Paris-Saclay, CEA, CNRS, SPEC, 91191 Gif-sur-Yvette, France}

\date{\today}
\begin{abstract}
Superconducting boron-doped silicon is a promising material for integrated silicon quantum devices.
In particular, its low electronic density and moderate disorder make it a suitable candidate for the fabrication of large inductances with low losses at microwave frequencies.
We study experimentally the electrodynamics of superconducting silicon thin layers using coplanar waveguide resonators, focusing on the kinetic inductance, the internal losses, and the variation of these quantities with the resonator read-out power.
We report the first observation in a doped semiconductor of microwave resonances with internal quality factors of a few thousand.
As expected in the BCS framework, superconducting silicon presents a large sheet kinetic inductance in the 50-500 pH range, comparable to strongly disordered superconductors.
The kinetic inductance temperature dependence is well described by Mattis-Bardeen theory.
However, we find an unexpectedly strong non-linearity of the complex surface impedance that cannot be satisfyingly explained either by depairing or by quasiparticle heating.
\end{abstract}
\maketitle

\section{Introduction}
Lossless high-inductance microwave components made from superconducting materials are currently under intense study, both theoretically and experimentally, as they play a crucial role in cavity QED~\cite{Samkharadze2016, Kuzmin2019, Leger2019}, in protecting qubits from their electromagnetic environment~\cite{Manucharyan2009}, in fast two-qubit gates~\cite{Harvey2018}, and in sensitive astronomical detectors~\cite{Day2003, Barends2007, Zmuidzinas2012}.
The development of such components made from superconducting silicon would be of particular interest for silicon-based quantum electronics, initiated with the demonstration of CMOS silicon spin qubits~\cite{Maurand2016}.
Silicon becomes a superconductor when a few percent of the silicon atoms are replaced by boron~\cite{Bustarret2006}. 
Using a pulsed laser doping technique, layers of superconducting silicon ultra-doped with boron (Si:B) can be fabricated with controlled thickness and dopant concentration. This control allows, in turn, to tune the superconducting critical temperature $T_\mathrm{c}$ in the \SI{0}{K} - \SI{0.7}{K} range \cite{Marcenat2010, Bhaduri2012, Grockowiak2013}.
However, while the DC properties of such tunable BCS superconductor are now well understood, little is known about its surface impedance at the microwave frequencies that are relevant for quantum information applications.
The surface impedance of thin-film superconductors is often dominated by the kinetic inductance $L_\mathrm{k}$, due to the inertia of the Cooper-pair condensate.
Indeed, $L_\mathrm{k}$ is proportional to the ratio of the normal-state sheet resistance $R_\square$ just above $T_c$ and the superconducting gap energy $\Delta$.
Disordered superconductors and very thin films of elemental superconductors may then have such high $R_\square$ that the kinetic inductance strongly exceeds the "geometrical" inductance due to energy storage in the magnetic field.

Here, we report the first observation in a superconducting doped semiconductor of microwave resonances with internal quality factors of a few thousand.
We show that the kinetic inductance of our thin Si:B films is large, and comparable to strongly disordered superconductors such as TiN~\cite{Swenson2013,Leduc2010,Shearrow2018}, NbN~\cite{Annunziata2010,Niepce2019,Gao2007}, NbTi~\cite{McCambridge1995}, NbSi~\cite{Calvo2014}, W~\cite{Basset2019} and granular Al~\cite{Maleeva2018}.
Si:B kinetic inductance measurements match the predictions based on DC measurements of $T_\mathrm{c}$ and $R_\square$, where $R_\square$/$T_\mathrm{c}$ is large due to the low carrier density and moderate disorder in the doped semiconductor.
We show that the temperature and frequency dependence of the kinetic inductance also follow the predictions of the BCS-based Mattis-Bardeen equations.
However, the non-linearity, manifested by the increases in both kinetic inductance and internal dissipation with increasing read-out power, is unexpectedly large, and cannot be fully explained as either depairing~\cite{Semenov2016} or quasiparticle heating~\cite{Guruswamy2015}.

\begin{figure*}[t!]
\includegraphics[width=\textwidth]{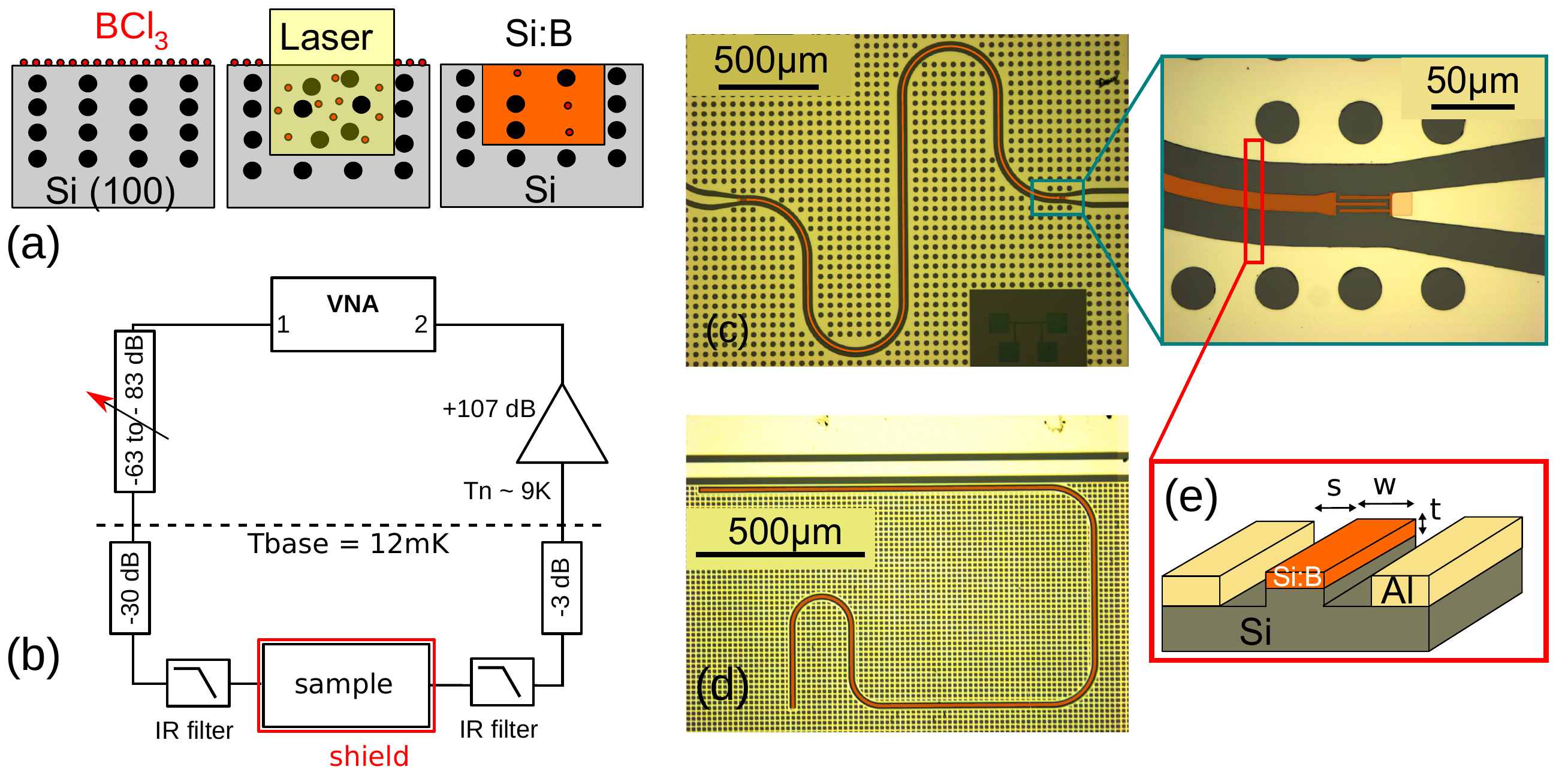}
\caption{
Fabrication, measurement and design of Si:B resonators.
\textbf{a} The steps in the Gas Immersion Laser Doping process, from left to right: chemisorption of the precursor gas BCl$_3$ onto the Si surface under ultra-high vacuum conditions; laser melting of the substrate by a \SI{25}{ns} excimer laser pulse and boron diffusion throughout the liquid phase; fast epitaxy of a superconducting Si:B crystal on the Si substrate.
\textbf{b} Simplified cryogenic microwave setup.
The sample is enclosed in a box made of either Au-plated Cu or superconducting Al, which, with a 10 mK shield, provides electromagnetic protection.
The input line, on the left, is heavily attenuated and filtered to be able to probe the resonators down to the single photon regime.
At the base temperature stage, we place homemade IR filters made of Eccosorb on both the sample input and output, as well as either an attenuator (as shown here) or a \SIrange{3}{12}{GHz} isolator on the output line to block thermal radiation from the input of the first stage HEMT amplifier at \SI{4}{K}.
\textbf{c,d} Optical micrographs of the two geometries of CPW resonators used in this work.
The Si:B sections are colorized in orange. 
\textbf{c} Sample F, transmission type.
\textbf{d} Sample C, shunt-coupled type.
\textbf{e} Cross-section of the CPW resonators.
In both geometries, the connecting leads and ground plane are made of Al, and the CPW central conductor is patterned from the Si:B layer, leaving the Si substrate in the gap between Al and Si:B.}
\label{fig_GILD_setup_sample}
\end{figure*}

\section{Epitaxial superconducting silicon}
Silicon strongly doped with boron is a type II disordered BCS superconductor~\cite{Bustarret2006} with a critical temperature that increases with increasing boron concentration up to \SI{0.7}{K}~\cite{Grockowiak2013,Dahlem2010}.
A nonequilibrium doping technique is necessary to create a superconductor because the minimal boron concentration required $n_\mathrm{B} \sim \SI{3e20}{cm^{-3}}$~\cite{Grockowiak2013} is about three times the solubility limit~\cite{Murrell2000}.
For this purpose, we use gas immersion laser doping (GILD)~\cite{Debarre2002}, whose principle is shown in Fig.~\ref{fig_GILD_setup_sample}a and in Methods.
The GILD technique turns the silicon substrate, over a thickness \SIrange{5}{300}{nm}, into epitaxial layers with a tunable boron concentration up to 10~at.$\%$ (\SI{5e21}{cm^{-3}}), without the formation of boron aggregates~\cite{Hoummada2012}.
The layers are homogeneously doped in thickness except for a few nanometer metallic layer at the interface between the Si and Si:B, over which the doping drops to 0.
The laser intensity has spatial inhomogeneities $\sim \SI{1.5}{\percent}$, which result in spatial fluctuations of the layer thickness, and may also result in small fluctuations of the critical temperature.

The resonators described in this work are patterned from Si:B layers epitaxied on a high-resistivity ($\rho > \SI{3}{k\ohm.cm}$) Si substrate.
These layers have $\SI{2}{mm} \times \SI{2}{mm}$ surface area, thickness $t$ from \SIrange{23}{35}{nm} and boron concentration 4 to \SI{4.7e21}{cm^{-3}} (8 to 9.5~at.$\%$).
The layers have a critical temperature $T_c$ from \SIrange{200}{500}{mK}, a maximum transition width $\Delta T_c = \SI{50}{mK}$ and a sheet resistance $R_\square$ from \SIrange{30}{80}{\ohm}, corresponding to resistivities \SIrange{1}{2}{\times ~10^{-4}~\ohm.cm}.
The layer thickness is much larger than the electronic mean free path $l_e \sim \SI{5}{nm}$, and is comparable or smaller than the superconducting coherence length $\xi \sim $ \SIrange{50}{110}{nm}.
These parameters, for all the resonators investigated here (labeled from A to I), are summarized in Table~\ref{table_summary} together with their geometrical and electrical properties.

\begin{table*}[t]
\includegraphics[width=\textwidth]{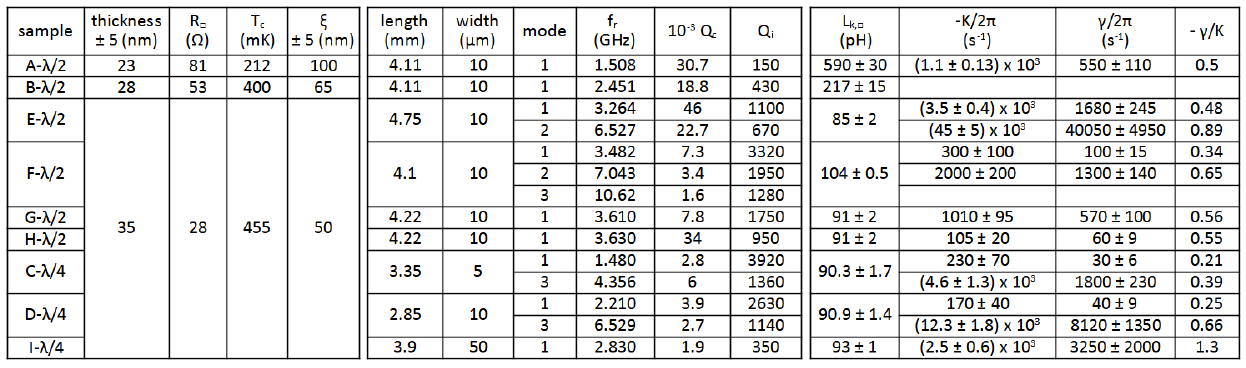}
\caption{
Summary of the geometric and electric properties of the measured samples: thickness $t$, sheet resistance $R_\square$, critical temperature $T_c$, superconducting coherence length $\xi$, length and width of the Si:B central conductor of the CPW resonators, resonance mode, resonant frequency $f_m$, coupling quality factor $Q_c$, internal quality factor $Q_{\mathrm{i}}$, sheet kinetic inductance $L_{k,\square}$, non-linear coefficients Kerr $K$ and $\gamma$, and their ratio (see text for details).
}
\label{table_summary}
\end{table*}

\section{Silicon resonator design and fabrication}
Our superconducting coplanar waveguide (CPW) resonators are fabricated as follows: first, the central conductor is patterned in a Si:B layer by optical lithography and a fluorine-based reactive ion etch; then, the ground planes and leads are patterned via lift-off from a \SI{200}{nm} evaporated Al film.
The ground plane thickness is comparable to the etched depth around the Si:B line.
The use of Al allows (i) keeping a constant impedance matching of our samples to the microwave equipment regardless of the Si:B surface impedance, and (ii) simplifying the system under study to a single conductor, allowing a more accurate extraction of the kinetic inductance value~\cite{Gao2008}.
The Al ground plane in the vicinity of the resonator contains a grid of \SI{10}{\micro\meter} wide holes acting as magnetic flux traps to limit the dissipation due to vortex motion~\cite{Stan2004} (Fig.~\ref{fig_GILD_setup_sample}c,d).

We characterize both half-wave ($\lambda/2$) and quarter-wave ($\lambda/4$) resonators.
The half-wave resonators are coupled at both ends to on-chip \SI{50}{\ohm} Al transmission lines via small capacitances.
These resonators are measured in a transmission configuration: the measured signal propagates through the resonator (Fig. \ref{fig_GILD_setup_sample}c).
The quarter-wave resonators are coupled at their open end to a \SI{50}{\ohm} transmission line, while their opposite end is shorted to the Al ground plane.
These resonators are measured in a shunt configuration, meaning that the measured signal is the complex sum of the signal propagating past the resonator and the signal radiated from it in the forward direction (Fig.~\ref{fig_GILD_setup_sample}d).
The latter geometry is best suited for reading out several resonators with different resonance frequencies in a single run.
In our samples, the transmission line crosses three laser-doped spots each containing one resonator.
Four-wire DC measurements were performed in parallel on laser doped samples fabricated at the same time.

The design of the resonators was determined by adjusting with microwave simulations in SONNET software the resonance frequencies $f_r$ and coupling loss rates $\kappa_c$ (see Methods).
To achieve an accurate determination of the resonators' internal loss rates $\kappa_\mathrm{i}$, we target values of the coupling losses $\kappa_c$ near the critical coupling $\kappa_c \sim \kappa_\mathrm{i}$.
$\kappa_c$ is in the range $3 - 150 \times 10^5 $s$^{-1} $, corresponding to quality factors  $Q_{c} = 2 \pi f_r / \kappa_{c} \sim $ 2000 to 46000, while $\kappa_\mathrm{i} \sim 20 - 600 \times 10^5 $s$^{-1}$ corresponds to quality factors $Q_{\mathrm{i}} = 2 \pi f_r / \kappa_{\mathrm{i}} \sim$ 150 to 4000.
(See Table~\ref{table_summary}).

\section{Measurement setup and linear response}
We use conventional microwave spectroscopy measurements of the superconducting Si:B resonators to determine the kinetic inductance of the film, the resonator internal loss rates $\kappa_\mathrm{i}$ and the amount of non-linearity in both of these quantities.
We extract these values from measurements of the complex forward scattering parameter $S_{21}$ at frequencies from \SIrange{1}{12}{GHz}, at temperatures from \SIrange{12}{350}{mK}, and using power levels $P_{in}$ at the sample input that range from \SIrange{-145}{-90}{dBm}.
At resonance, the photon number in the resonator is 
$\overline{n} = P_{in} \nicefrac{2 \kappa_c}{\hbar \omega_r \kappa_t^2}$  where $\kappa_t = \kappa_c + \kappa_\mathrm{i}$ is the total loss rate, and varies from less than 1 to a few million for the power levels used here.
As shown in Fig.~\ref{fig_GILD_setup_sample}b, both input and output lines are shielded to prevent significantly populating the resonators with thermal photons.

The measured $S_{21}$ is the product of the signal of interest transmitted through the chip and the background transmission due to the input and output lines. 
In the shunt configuration, the background transmission can be measured in the normal state of Si:B, while the Al is in the superconducting state.
This compromise is best achieved around \SI{400}{mK}, where no Si:B resonance can be detected, and where the background transmission is almost independent of the temperature.
In the transmission configuration, the background was estimated in a separate run.
With this background removed, the transmission coefficient used to fit data in the shunt configuration is
\begin{equation}
S_{21} = 1 - (1 + \mathrm{i}A) \frac{\kappa_c}{\kappa_c + \kappa_\mathrm{i} + 2 i \delta \omega},
\label{s21}
\end{equation}
while in the transmission configuration it is
\begin{equation}
S_{21} = S_L + \frac{2 \kappa_c}{2 \kappa_c + \kappa_\mathrm{i} + 2 \mathrm{i} \delta \omega},
\label{s21t}
\end{equation}
where $\delta \omega = 2 \pi (f - f_r)$ is the detuning.
The asymmetry $A$ and leakage $S_L$ account for on-chip imperfections: $A$ is nonzero when the field radiated from the resonator reflects off an impedance mismatch such as that occurring at the micro-bonding connections~\cite{Khalil2012,Megrant2012}, while $S_L$ is nonzero when the input signal couples directly to the output.

\begin{figure*}[!t]
\includegraphics[width=\textwidth]{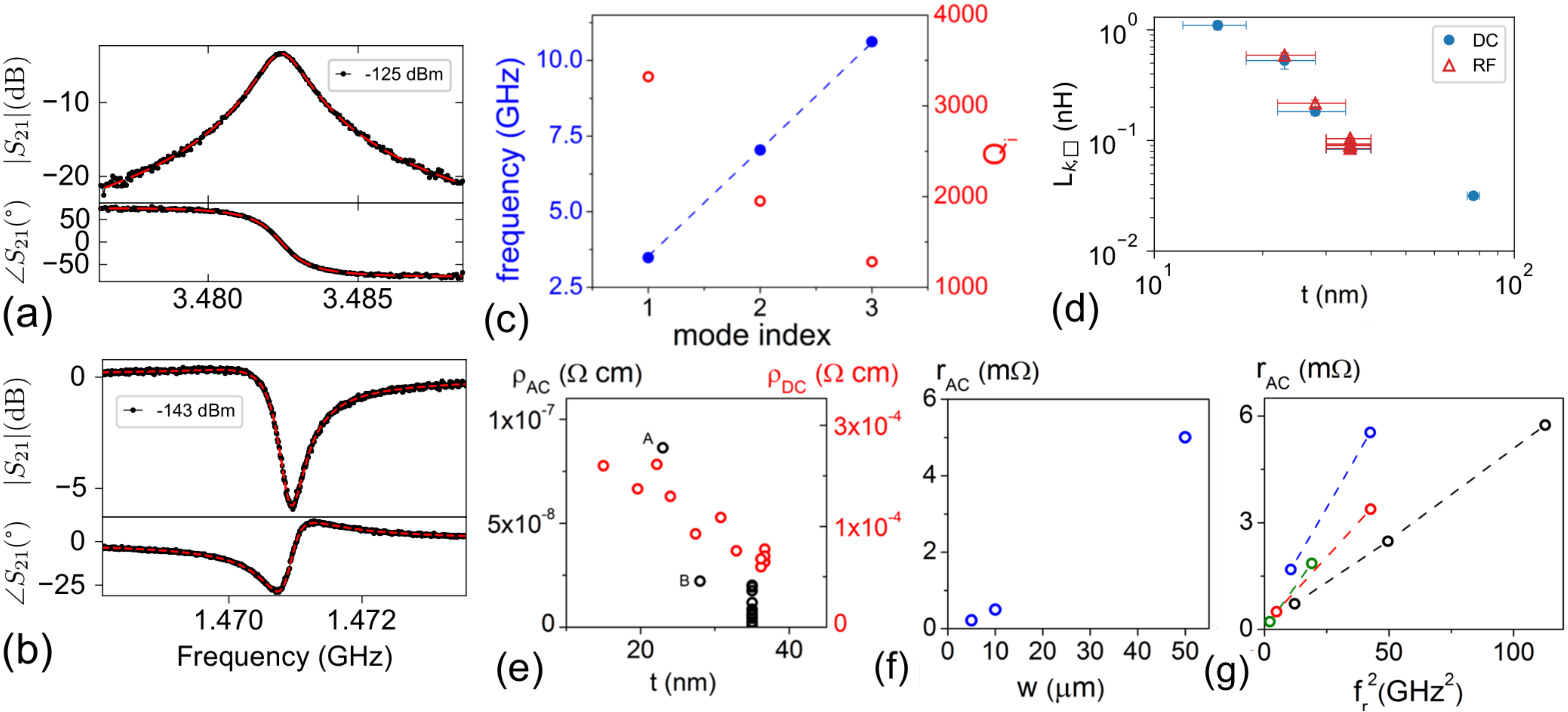}
\caption{ \textbf{Linear regime} - 
\textbf{a} Typical resonance curve measurement (black dots) in the transmission-coupled geometry with the setup background transmission removed.
The data is taken from sample F at temperature $T = \SI{14}{mK}$ and read-out power $P = \SI{-125}{dBm}$.
The red dashed line is a fit of Eq.~(\ref{s21t}) to the data.
\textbf{b} Typical resonance curve measurement for shunt-coupled sample C, with $T = \SI{11}{mK}$, $P = \SI{-143}{dBm}$, and the setup background transmission removed.
The red dashed line is a fit of Eq.~(\ref{s21}).
The curve distortion is not related to non-linear effects, but is due to impedance mismatch, modelled by the asymmetry parameter $A$ in Eq.~(\ref{s21}).
\textbf{c} Resonance frequency and quality factor of the first three modes of sample F (half wave) measured at $T = \SI{14}{mK}$.
The black dashed line follows $f_m = m * f_1 $, where $m$ is the mode index.
The internal quality factor is always found to decrease with increasing frequency for a given device.
\textbf{d-g} \textbf{Kinetic inductance and losses in the linear regime}. Summary of the main RF figures gathered in Table~\ref{table_summary} for all samples measured in this work, labelled when needed.
\textbf{d} Sheet kinetic inductance $L_{k,\square}$ as a function of Si:B thickness inferred from measurements at DC (blue dots) or at RF (red triangles).
The error bars attributed to the thickness are associated to the uncertainty of the thickness of the superconducting layer within the doped layer.
\textbf{e} The AC sheet resistivity $\rho_{AC}$ is found to decrease with increasing thickness. \textbf{f} the AC sheet resistance $r_{AC}$ increases with the Si:B central conductor width, and \textbf{g} increases with the square of the frequency.}
\label{fig_summary_linear}
\end{figure*}

The linear response is measured with a minimum applied power corresponding to a few photons on average in the resonator, where the curve $S_{21}(f)$ is independent of the read-out power.
The experimental data are fitted with Eq.~(\ref{s21}) or Eq.~(\ref{s21t}) up to \SI{350}{mK}, where most resonances vanish in the noise.
From these fits we extract $f_r$, $\kappa_\mathrm{i}$ and $\kappa_c$ as a function of temperature up to $T \sim 0.6 \, T_c$.
Characteristic measurements with the corresponding fits are presented in Figs.~\ref{fig_summary_linear}a and b.
As expected, our resonators are multi-modal, with resonant frequencies scaling with the mode number $m$ as $f_m = m f_1$ for half-wave resonators and as $f_m=(2m-1) f_1$ for quarter-wave resonators.
In our experimentally accessible frequency range we can thus observe the first two or three modes of each resonance.

\section{Low temperature kinetic inductance in the linear regime}
In the BCS theory, the kinetic inductance at low temperature ($T \ll T_c$) and low frequency ($hf \ll k_\mathrm{B} T_c$) is:
\begin{equation}
L_{k} = \frac{\hbar R_\square}{\pi \Delta} \, \frac{1}{\tanh (\Delta/2 k_\mathrm{B} T)}.
\label{eqLk}
\end{equation}
In the case where the inductance is so high that the current density is uniform, it is convenient to introduce the sheet kinetic inductance $L_{k,\square}$, as is done for the sheet resistance.
This quantity is relevant in our experiments, where only the CPW resonator central conductor is made from Si:B, and has a small cross-section.

Using Eq.~\ref{eqLk} with the sheet resistance $R_{N,\square}$ and $T_c$ extracted from DC measurements,
$L_{k,\square}$ is found to be in the range \SIrange{50}{470}{\pico\henry\per\sq} for Si:B thicknesses ranging from \SIrange{20}{80}{nm}. 
These values are then compared to those extracted from the RF measurements by adjusting the SONNET electromagnetic simulations to reproduce the measured resonance frequencies.
Note that the $L_{k,\square}$ values deduced from the SONNET simulations are found in good agreement with the analytical values obtained from the resonant frequency
$f_1 = 1/p l \sqrt{\mathcal{C}_g (\mathcal{L}_g+\mathcal{L}_k})$, for the first mode, where $l$ is the length of the resonator, $p = 2 (4)$ for a $\lambda/2$ ($\lambda/4$) resonator, and $\mathcal{C}_g$, $\mathcal{L}_g$, and $\mathcal{L}_k$ are respectively the capacitance, geometric inductance and kinetic inductance per unit length. This expression, where the current distribution (relatively homogeneous) in the width of the central conductor was taken into account (see Methods), is valid when the losses can be neglected.
Fig.~\ref{fig_summary_linear}d shows excellent agreement between the $L_{k,\square}$ values inferred from the high frequency measurements and those calculated from the DC characteristics, confirming the validity of the BCS description and suggesting that values of the order of \SI{1}{\nano\henry\per\sq} could be obtained in Si:B resonators patterned from thinner layers. 
For all our samples, the kinetic inductance participation ratio $\alpha=L_k/(L_k+L_g)$ ranges from 0.94 to 0.99.
The resonator characteristic impedance $Z_r=\sqrt{\mathcal{L}_{tot}/\mathcal{C}_g}$ lies in the range \SIrange{150}{700}{\ohm}.
In particular, the \SI{35}{nm} thick resonators, whose results are mainly shown here, have $Z_r \sim \SIrange[range-phrase=-, range-units=brackets]{300}{380}{\ohm}$.
\begin{figure*}[t!]
\includegraphics[width=\textwidth]{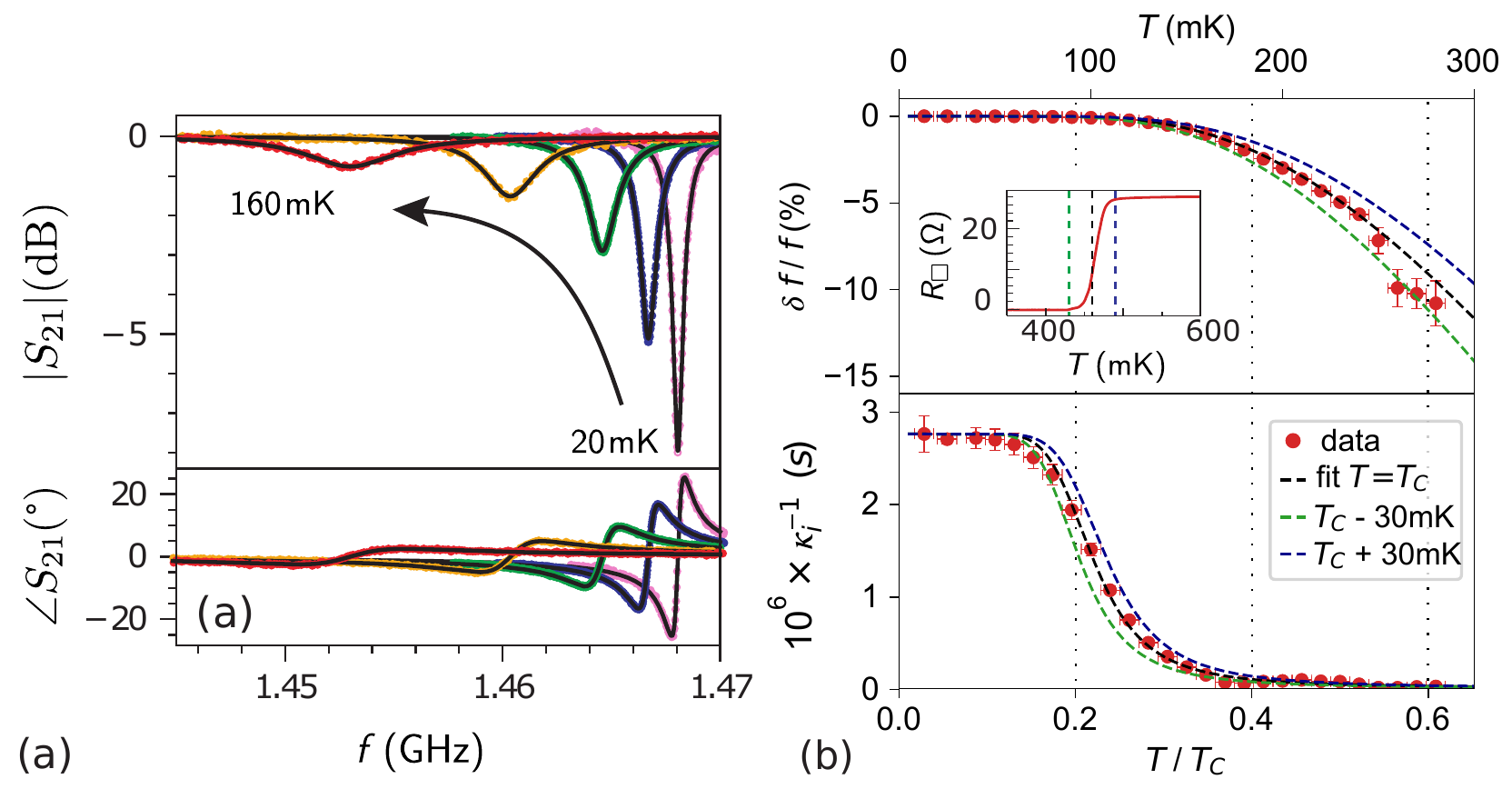}
\caption{
Evolution of one resonance with temperature for sample C obtained in several cool-downs, and comparison with the Mattis Bardeen theory.
\textbf{a} $S_{21}(f)$ data (dots) from \SIrange{20}{160}{mK} after removal of the reference background taken at \SI{400}{mK}, compared with fits of Eq.~(\ref{s21}) (black lines).
\textbf{b} Relative frequency shift and inverse internal losses as a function of temperature expressed in reduced units of $T_c$.
The black dashed line is the Mattis-Bardeen prediction using the measured $T_c$, with no free parameters.
Colored dashed lines are fits using $T_c \pm \SI{30}{mK}$.
\textbf{b}-inset Four-wire DC measurements of the sheet resistance $R_\square$ of the same Si:B layer, showing $T_c = \SI{460 \pm 30}{mK}$ and $R_\square = \SI{28}{\Omega}$ in the normal state just above $T_c$.
}
\label{fig_MattisB}
\end{figure*}

\section{Low temperature losses in the linear regime}
We evaluate the residual resistance at microwave frequency in the Si:B layer with a resonator effective RLC lumped-element model.
As for the extraction of the sheet kinetic inductance from the resonance frequency, we aim at extracting the residual sheet resistance from the losses, describing our material independently of the device geometry. 
We compute the AC sheet resistance from the internal quality factor and sheet kinetic inductance with $r_{AC} = \omega_r \, L_{k,\square} /\alpha \, Q_\mathrm{i,sat} $, with $Q_\mathrm{i,sat}$ the base temperature, saturated, internal quality factor of the mode.
This sheet resistance is reflective of the ohmic losses in the inductor and ranges from \SIrange{0.2}{40}{\milli\ohm\per\sq}.
We find that the AC losses depend on the layer thickness, as already observed for the DC resistivity (Fig.~\ref{fig_summary_linear}e): when the thickness decreases, both AC and DC Si:B resistivities ($\rho_{AC}=r_{AC} \,t$; $\rho_{DC}=R_\square \,t$) increase as a result of a larger amount of interstitial boron and, correspondingly, of disorder~\cite{Bonnet_th_2019} (see Methods).
The highest $r_{AC}$ are observed for the thinner, lossier, more disordered samples, which present the highest $L_{k,\square}$. 
Furthermore, $r_{AC}$ strongly increases with the resonator width, as observed in Fig.~\ref{fig_summary_linear}f for samples C, D and I, whose laser doping and fabrication were realised at the same time, and that were probed by the same transmission line. 
Finally, $r_{AC}$ is plotted against frequency in Fig.~\ref{fig_summary_linear}g, for the samples where multiple harmonics could be measured, and was shown to increase as $f_r^2$. 
The frequency and width dependencies are in quantitative agreement with AC losses due to magnetic vortices, for which the motion of the non-superconducting vortex core leads to dissipation (see Methods). However, in a control experiment with extensive magnetic shielding on sample C the losses were not found to vary significantly. This could indicate that another type of vortex loss involving vortex-antivortex pairs is at play.
Other features can also contribute to the observed losses.
For instance, a few nanometer thick metallic layer is present at the Si:B/Si interface, where the dopant concentration decreases below the superconductivity threshold, and whose thickness depends on the layer properties. However the associated AC sheet resistance is not expected to depend on the conductor width.
Further experiments are still needed to discriminate between the possible sources of loss.

\section{Temperature dependence of resonant frequency and losses - Mattis-Bardeen theory}
Fig.~\ref{fig_MattisB}a shows a selection of the background-removed $S_{21}$ data measured in the linear response regime at several temperatures for sample C.
A reduction of the resonant frequency and quality factor is observed, as the thermally-induced Cooper pair breaking increases the kinetic inductance and the losses~\cite{Gao2008}.
In Fig.~\ref{fig_MattisB}b, we compare the characteristic temperature dependence of the relative frequency shift $\delta f / f $ and inverse internal losses $\kappa_\mathrm{i}^{-1}=Q_\mathrm{i}/\omega$ to the prediction of Mattis-Bardeen theory~\cite{Mattis1958}, valid for homogeneous weak-coupled BCS superconductors with low or moderate disorder.
The evaluation of the frequency-dependent complex conductivity is based on two input parameters extracted from DC measurements: $T_c$ and $R_\square$.
Fig.~\ref{fig_MattisB}b shows the good agreement obtained for the frequency shift, without any free parameters, up to temperatures as high as $0.6 \,T_c$ where $\delta f/f$ decreases by 10$\%$.
We add to Fig.~\ref{fig_MattisB}b, to guide the eye, the Mattis-Bardeen expectations for $T_c + \SI{30}{mK}$ and $T_c - \SI{30}{mK}$.
Such temperature range roughly corresponds to the width of the superconducting transition of the sample, as shown in the inset of Fig.~\ref{fig_MattisB}b.
Above $0.6\, T_c$ the determination of $\delta f/f$ and $\kappa_\mathrm{i}$ becomes increasingly difficult due to the vanishing amplitude of the field coming out of the resonator (since $\kappa_\mathrm{c} << \kappa_\mathrm{i}$), and to the low read-out power required to remain in the linear regime.
Despite the increased uncertainty, however, a small shift from the expected dependence can be observed above $0.55 \, T_c$.
This may be the result of some inhomogeneities in $\Delta$ in the Si:B layer, as a gap dispersion of the order of $\pm 10 \,\%$ was measured by scanning tunneling spectroscopy~\cite{Dahlem2010}.
\begin{figure*}[t!]
\includegraphics[width=\textwidth]{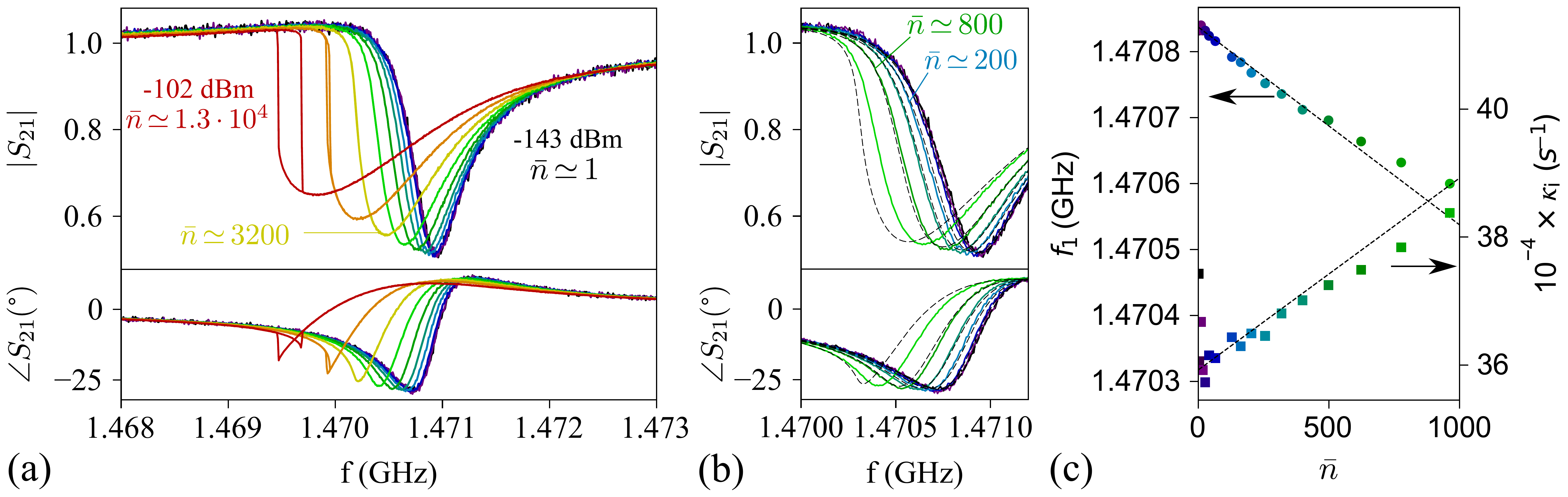}
\caption{
Measurement and fit of the non-linear scattering parameters of sample C. 
\textbf{a} Evolution of the resonance with increasing input power for the first mode at $T = \SI{14}{mK}$.
\textbf{b} A low-power subset of measured $S_{21}$ is fitted with Eq.~(\ref{s21_NL}) (dashed lines) for a common value of $K_{low}/2\pi = \SI{-300}{s^{-1}}$ and  $\gamma_{low}/2 \pi = \SI{30}{s^{-1}}$.
\textbf{c} For the same subset, apparent resonance frequency (dots) and internal losses (squares) are deduced from the position and amplitude of the dip in $|S_{21}(f)|$, and plotted against average photon number $\overline{n}$.
The symbol color encodes the input power with the same scale as in insets \textbf{a} and \textbf{b}.
Dashed lines correspond to the same $K$ and $\gamma$ of the complete fits: $\partial f_1 / \partial \overline{n} = K_{low}/2\pi$ and $\partial \kappa_\mathrm{i} / \partial \overline{n} = \gamma_{low}$.
}
\label{fig_NLs21_KerGamma}
\end{figure*}

Fig.~\ref{fig_MattisB}b shows the temperature dependence of the internal losses. At high temperature ($T > 0.3\, T_c$), losses due to thermally activated quasiparticles are well-described by Mattis-Bardeen ($\kappa_\mathrm{i} \sim \kappa_\mathrm{MB}$).
At the base temperature, $\kappa_\mathrm{i}$ is found to saturate due to additional loss mechanisms, giving a saturation value $\kappa_\mathrm{sat}$.
The total internal losses are written $\kappa_\mathrm{i} (T)=\kappa_\mathrm{MB} (T)+\kappa_\mathrm{sat}$.
This saturation value takes into account all the usual non-thermal contributions, such as quasiparticles excited by unshielded radiation or by the read-out power, the dissipation due to two-level systems or by vortex motion, etc.
In conclusion, the linear response of Si:B at finite frequency shows a good agreement with the BCS-based Mattis-Bardeen theory, at all temperatures for the purely reactive (inductive) response.
At temperatures $T < 0.2 \,T_c$ however, a loss mechanism, discussed when commenting the AC sheet resistance at the base temperature, takes over the dissipative response.

\section{Non-linearities of kinetic inductance and losses}
The results shown up to now are characteristic of the linear regime. However, when increasing the readout power, the resonances shift towards the low frequencies, then become progressively skewed, up to a bistable state where an hysteresis is observed between frequency sweeps up and down (Fig.~\ref{fig_NLs21_KerGamma}a). 
Alongside with this change of resonance frequency and shape, we also note an increase in the internal losses, with a reduction of the resonance amplitude. 
From a first qualitative point of view, in Si:B, this non-linearity appears at extremely low power: as an example, the first mode ($f_1 = \SI{1.48}{GHz}$) of sample C undergoes frequency shifts of the order of \SI{100}{kHz} at read-out powers of the order of fW, corresponding to about 200 photons in the resonator, while the hysteresis appears for pW powers.
To describe quantitatively the resonance curves in the non-linear regime, the harmonic oscillator Hamiltonian characteristic of the linear regime is replaced by $H=\hbar \omega_r a^{\dagger}a + \hbar/2  K a^{\dagger} a^{\dagger} a a$,
where $a^{\dagger}$ and $a$ are the photon creation and annihilation operators, and $K$ is the Kerr coefficient, quantifying the non-linear response.
The energies of this oscillator are no longer equidistant, but acquire a Fock state dependent shift
$(E_{n+1} - E_{n})/ \hbar = \omega_r + n K $,
where $n$ is the Fock index.
In consequence, the measured resonance shifts with the average photon population $\overline{n}$ writes $\omega_r^{NL} = \omega_r + \overline{n} K $ \cite{Yurke2006}.
Non-linear losses are also taken into account, introducing a loss term proportional to the number of photons and writing the total losses $\kappa_{tot}^{NL} = \kappa_c + \kappa_\mathrm{i} + \overline{n} \gamma $.
The transmission coefficient, in the case of a shunt-coupled geometry finally reads
\begin{equation}
S_{21}^{NL}= 1 - \frac{(1+iA) \, \kappa_c}{(\kappa_{c}+\kappa_\mathrm{i} + \overline{n} \gamma) + 2 i [\omega -(\omega_r + \overline{n} K)]}
\label{s21_NL}
\end{equation}
A similar expression is found in the transmission geometry.
Fig.\ref{fig_NLs21_KerGamma}b shows the results of the fits from Eq.~(\ref{s21_NL}), where the curves in the low power range, for both the up and down sweeps, are described by a single Kerr number $K$ and a single non-linear loss term $\gamma$.
The other parameters in the fit, $\kappa_\mathrm{c}$, $\kappa_\mathrm{i}$, $f_r$, and the asymmetry $A$, are fixed at the linear regime values.
All the curves measured in the wider power range, corresponding to $\overline{n}=1-10^4$ photons in the resonators, are well-described by Eq.~(\ref{s21_NL}), but only the low-power curves, with $\overline{n} \lesssim 1000$ can be described by a common $K$ and $\gamma$. 
This deviation may be due to higher order terms in the Hamiltonian.
Henceforth, we will exclusively discuss the low-power non-linear parameters. 
Note that the low-power Kerr values and non-linear losses deduced from the non-linear fit are in good agreement, within 20$\%$, with the linear variation with $\overline{n}$ of the resonant frequency and losses extracted from linear fits of $|S_{21}(f)|$ for $\overline{n} \lesssim 500$ (see Fig.~\ref{fig_NLs21_KerGamma}c). 

The extracted $K/2\pi$ and $\gamma/2\pi$ are reported in Table~\ref{table_summary} for the measured harmonics of all resonators.
The fits output a negative $K$ and a positive $\gamma$, consistent with the expectation that, whatever the non-linearity mechanism, it corresponds to a weakening of superconductivity and thus an increase of the kinetic inductance and additional losses in the system.
The smallest Kerr numbers reported are, in absolute value, of the order of \SI{100}{Hz/photon} for macroscopic CPW structures, showing the extreme sensitivity of superconducting silicon to microwave power.
\begin{figure*}[t!]
\includegraphics[width=\textwidth]{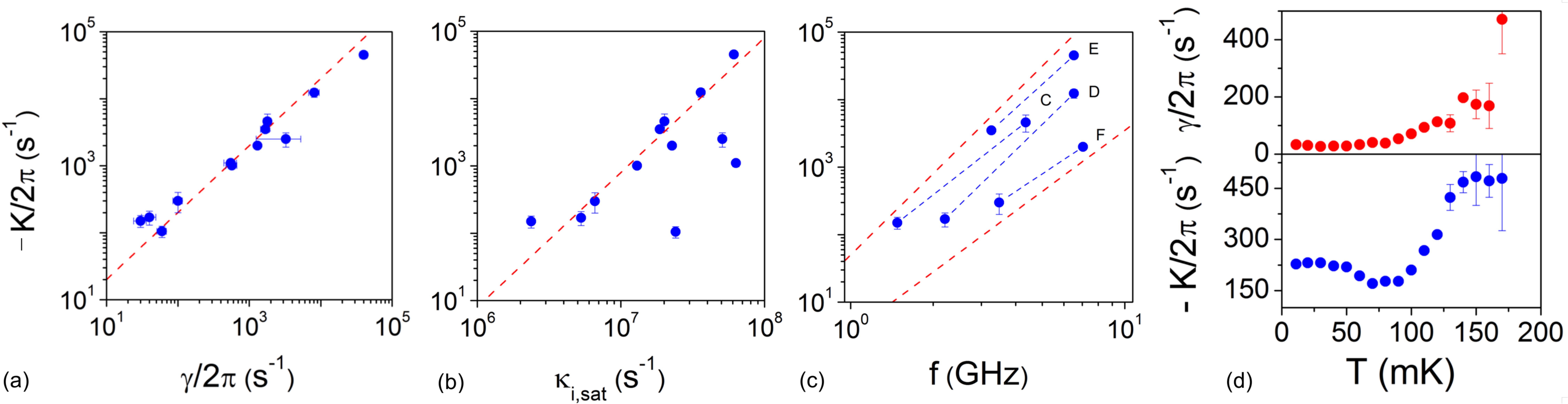}
\caption{
General trends for the non-linearity in Si:B. 
\textbf{a} reactive vs dissipative non-linearity for all the samples reported in Table~\ref{table_summary} : Kerr coefficient vs. non-linear losses $\gamma$.
The dashed line corresponds to $\gamma / K = 0.5$.
\textbf{b} Kerr coefficient vs. linear losses at the base temperature (20 mK)  for all the samples reported in Table~\ref{table_summary} .
The dashed line corresponds to $K \propto \kappa_{i,sat}^2$.
\textbf{c} Kerr coefficient vs. frequency for samples where the non-linearity was measured for multiple harmonics.
The dashed lines (shifted for clarity) correspond to $K \propto f^3$ and $K \propto f^4$.
\textbf{d} Kerr coefficient and non-linear losses vs. temperature for sample C.
Labels correspond to Table~\ref{table_summary}.
}
\label{fig_NL_summary}
\end{figure*}
In Fig.~\ref{fig_NL_summary} we plot the variation of the Kerr coefficient against the non-linear losses $\gamma$, the saturation losses in the linear regime $\kappa_{i,sat}$, the frequency and the temperature.
The non-linear inductive response ($K$) is associated to non-linear losses ($\gamma$) over a broad range of magnitudes (Fig.~\ref{fig_NL_summary}a), with a ratio $\gamma/ K \sim 0.5$ (dashed line and Table \ref{table_summary}).
At the base temperature, the non-linearity is higher for the samples presenting higher losses in the linear regime (Fig.~\ref{fig_NL_summary}b).
$K$ and $\gamma$ strongly depend on frequency, with $K \sim f^{3} - f^4$ (dotted lines in Fig.~\ref{fig_NL_summary}c) and $\gamma \sim f^{3.5} - f^5$, with $K/\gamma \sim f$.
Finally, the temperature dependence of $K$ and $\gamma$ is measured for two samples (C, shown in Fig.~\ref{fig_NL_summary}d, and F).
The non-linearity globally increases with temperature.
To explain the characterised non-linearity, we examine in detail the two most common sources of non-linearity in superconducting resonators: the non-linearity of the kinetic inductance induced by depairing, and the non-linearity induced by quasiparticle heating.  

\section{Modeling non-linearity with depairing}
A non-linear behaviour with RF power is generally expected in superconducting resonators due to the non-linearity of the kinetic inductance with the current, $L_k = L_{k,0} (1+ I^2/I_{\Gamma}^2)$, where $I_{\Gamma}$, the depairing current, is of the order of the critical current ($I_\Gamma = 4.7 \times\, I_c$~\cite{Anthore2003}), and accounts for the weakening of superconductivity under a phase gradient.
Indeed, a static depairing (induced by a DC current, DC magnetic field or static magnetic impurities) induces a closing of the gap and smoothed BCS coherence peaks~\cite{Anthore2003}, while the density of states remains strictly zero below the reduced gap, like in the Abrikosov-Gorkov type impurity scattering.
In the case of a dynamic depairing (induced by an AC current or AC magnetic field), both the density of states and the distribution functions are modified, with in particular a non-zero density of states well below the gap~\cite{Semenov2016}.
To determine how much the conductivity deviates from the linear response at a certain current $I$, one has to evaluate the first order correction to the superconducting gap $\Delta(I)$, and hence to the kinetic inductance $L_k(I)$ in presence of depairing.
For a LC resonator, a relative shift of frequency is induced by a relative shift of inductance and is related to current following: 
\begin{equation}
\frac{\delta f}{f} = -\frac{\alpha}{2} \frac{\delta L_k}{L_k} = -\frac{\alpha}{2} \frac{I^2}{I_{\Gamma}^2} 
\label{}
\end{equation}

Expressing the energy stored in the inductor in terms of average number of photons
$\overline{n} \hbar \omega = 1/(2 \, \alpha) L_k I^2$
enables one to relate $K$ to the depairing current:
\begin{equation}
K_{dep} = \frac{2 \pi \delta f}{\overline{n}} \sim - \alpha^2 \frac{\hbar \omega^2}{L_k I_{\Gamma}^2}
\label{}
\end{equation}
To estimate the order of magnitude of $K_{dep}$, one can calculate $I_{\Gamma}$ from the measured critical current density in Si:B, $j_c \sim \SI{6e8}{A/m^2}$.
As an example, in the geometry of sample C, we obtain $|K_{dep}|/2\pi  \sim \SI{1}{Hz / photon}$, more than two orders of magnitude smaller than the  measured value of $|K_{mes}|/2\pi = \SI{150}{Hz / photon}$.
A more sophisticated approach that uses the Usadel equations has been shown to give accurate predictions of the Kerr coefficient for Al, NbSi, TiN and NbTiN resonators~\cite{Bourlet2021}:
\begin{equation}
K_{dep} = - \alpha^2 \frac{e^2}{h} \frac{9 \pi^2}{4} m^2 \frac{1}{\mathcal{C}_{g}} \frac{\xi^2}{l^3}
\label{K_hls}
\end{equation}
in a half-wave CPW superconducting resonator, where $m$ is the mode number, $\xi$ the superconducting coherence length and $l$ the length of the resonator.
Note that this prediction accounts for a static depairing, and not a dynamic depairing as described in Ref.~\cite{Semenov2016}.
However, it is justified in the limit $h f \ll \Delta$, where our measurements are performed.
Indeed, although the density of states is markedly different in the two cases, the conductivities are modified significantly by the dynamic depairing only when $h f$ is close to $\Delta$.
The estimations of $K_{dep}/2\pi$ based on Eq.~(\ref{K_hls}) show a discrepancy with the experimental values of about 2-3 orders of magnitude, although the scaling of $K_{dep}$ with frequency for multiple modes is reproduced. 
Moreover, the temperature dependence of $K_{dep}(T) \sim  \xi^2 \sim 1/\Delta(T)$, in the dirty limit where $\xi=\sqrt{\hbar D/\Delta}$, is observed neither in the shape nor in the magnitude
of $K(T)$ shown in Fig.~\ref{fig_NL_summary}d. 
Indeed, in both samples C and F, between \SI{11}{mK} and \SI{150}{mK}, $K$ increases by 100$\%$ and $\gamma$ by 370$\%$, while $\Delta$ varies only by 0.5$\%$.
Note that as seen from Eq.~(\ref{K_hls}), the only two varying parameters of the depairing model are $l$ and $\xi$, the CPW geometrical capacitance per unit length $\mathcal{C}_g \sim \SI{150}{pF/m}$ being mainly fixed by the substrate dielectric constant and $\alpha$ being near one for all our samples.
Both $l$ and $\xi$ are well known, the values of the superconducting coherence length being inferred from independent magneto-transport measurements for each Si:B layer~\cite{Bonnet_th_2019}, and reported in Table~\ref{table_summary}. 
The depairing model thus fails to predict the observed Kerr, which is dominated by another source of non-linearity.

\section{Modeling non-linearity with heating of quasiparticles}
Another possible explanation for the observed non-linearity is the heating of the quasiparticle (QP) system by the absorbed microwave power.
The quasiparticle system is cooled both by inelastic electron-phonon scattering (whose strength is proportional to the material dependent parameter $\Sigma_N$), as in a normal metal, and by quasiparticle recombination producing phonons with energies greater than $2 \Delta$ that may escape from the layer (whose strength is proportional to the material dependent parameter $\Sigma_S$) .
Whenever the microwave power absorbed by the QP system exceeds the cooling power due to electron-phonon scattering and quasiparticle recombination, the effective equilibrium temperature of the QPs increases, reducing both the resonance frequency and the internal quality factor (as seen in Fig.~\ref{fig_MattisB}).
The resulting equilibrium depends on the difference between the microwave frequency and the resonance frequency, as the maximum microwave absorption happens only at resonance.
As a consequence, heating leads to skewing and hysteresis of $S_{21}(f)$, as in the case of depairing, and could account for the experimental data shown in Fig.~\ref{fig_NLs21_KerGamma}a (see Methods). 
This model is detailed in~\cite{Goldie2012} for the normal metal electron-phonon cooling, and in~\cite{Guruswamy2015,Guruswamy_th_2018} for the superconducting case that we consider here.

Assuming the extreme scenario where all the missing power at the output port of the resonator is dissipated by QPs, we have computed the expected frequency shift as a function of the number of photons, deriving the 'apparent' Kerr and non-linear loss coefficients.

As ultra-doped silicon is a little-known material, we adjusted the value of $\Sigma_{S} \propto \Delta^4 \Sigma_\mathrm{N} $ ~\cite{Guruswamy_th_2018} to best reproduce the measured Kerr for each mode, assuming a thermal quasiparticle distribution. The obtained values are in the range $\Sigma_{S} = (0.02 - 20)$ nW µm$^{-3}$K$^{-1}$  with a mean around $\Sigma_{S} = \SI{1}{nW \micro m^{-3} K^{-1}}$, giving an unexpected large spread  for nominally identical samples, some of them even fabricated and measured simultaneously.
A theoretical lower limit for $\Sigma_S$ can be estimated from the value of $\Sigma_{N} = \SI{0.05}{nW \micro m^{-3} K^{-5}}$ of a Si:B film of smaller concentration \SI{5e20}{cm^{-3}} \cite{Chiodi2017}. We find
$\Sigma_{S} = \SI{0.02}{nW \micro m^{-3} K^{-1}}$, almost two orders of magnitude smaller than what we extract from most of our measurements. 
We would thus naively expect a much higher heating (and non-linearity) than what we observe. 
This implies that, at low temperature, we cannot rule out the contribution of heating to the non-linearity in Si:B.
However, both the Kerr amplitude and non-linear losses induced by heating are expected to decrease when the temperature increases, as the cooling becomes more efficient with the increase of the phonon population. This qualitative argument should hold also for a non-equilibrium, non-thermal distribution function with an excess of QPs \cite{deVisser-2014-prl}. However, the exact opposite is observed in Fig.~\ref{fig_NL_summary}d.
In addition, the expected ratio $\gamma_{th}/K_{th}$ for heating at equilibrium $\gamma_{th}/K_{th} = 0.85$ is approximately double of the measured one.
In conclusion, while we do not exclude quasiparticle heating at the lowest temperatures, another mechanism is dominant for higher temperatures above 100 mK.

\section{Other possible sources of non-linearity}
The correlation of linear and non-linear losses with the Kerr term suggest that the mechanism at the root of the non-linearity we observe in Si:B must comprise energy dissipation and heating of the system, and might also be responsible for the saturation of the internal losses in the linear regime.
In addition to a possible quasiparticle heating at the lowest temperatures, we suggest another non-linear mechanism inducing both a change of frequency and dissipation, related to 2D nature of the superconducting layer.
Indeed, as the Si:B resonator is thinner than the coherence length, vortex-antivortex bound pairs are present even in the absence of any magnetic field, pinned to the material defects.
We consider here the situation of our experiment where the induced AC magnetic field in the resonator is not strong enough to enable vortices to enter the superconductor.
These vortex-antivortex pairs, whose number and depinning increase with the amplitude of the AC bias, modify the current lines circulating in the superconductor, and can induce a strong non-linearity both in the reactive and in the dissipative response, as was shown for the extreme case of a granular superconductor modelled as a Josephson junction network~\cite{dahm2001}.
In our case, even though Si:B is not a granular material, the variations of the gap due to doping modulations may induce an array of Dayem weak links, qualitatively described by the same physics. When the temperature or the current increase, the bound states progressively dissociate, inducing a strong increase in the dissipation non-linearity.
To further test this hypothesis we plan to perform experiments in a magnetic field and with engineered flux traps in the CPW inner conductor, but such experiments are beyond the scope of the present work. 

\section{Conclusion}
In this work, we demonstrate the first resonating microwave circuits made of superconducting silicon.
They are realised from laser ultra-doped Si:B layers and standard micro-electronics processes.
Si:B presents a large kinetic inductance, in agreement with BCS expectations and in the range \SIrange{50}{470}{\pico\henry\per\sq}, comparable to strongly disordered superconductors. 
The temperature dependence of the kinetic inductance up to $0.6 \,T/T_c$ is similarly well understood in the framework of the Mattis-Bardeen theory, with no adjustable parameters.
The internal losses at high temperature ($T/T_c > 0.3$) are also well described by the Mattis-Bardeen theory as due to thermally-activated quasiparticles.
However, at low temperature the internal losses saturate, limiting the quality factor to a few thousand, possibly due to the contribution of vortex-induced dissipation. 
When increasing the read-out power, we evidence an unexpectedly strong non linearity in the kinetic inductance and the losses, two orders of magnitude larger than that predicted by depairing. The observed non-linearity increases with temperature, a dependence which is at odds with a quasiparticle heating model.
We speculate that this strong non-linearity is related to the 2D geometry of our superconducting silicon films.
This work opens the possibility of using Si:B in superconducting circuits on a quantum silicon platform, once its properties are optimised through a better understanding of the origins of the losses and of the large non-linearity.

\section*{Acknowledgements}

The authors are grateful to J. Basset, G. Hallais, I. Pop, P. de Visser, and K. Van der Beek for fruitful discussions.
FC acknowledges support from the R\'eseau RENATECH and from the French National Research Agency (ANR) under contract number ANR-16-CE24-0016-01, and as part of the “Investissements d’Avenir” program (Labex NanoSaclay, ANR-10-LABX-0035) 
HLS acknowledges support from the ANR ELODIS2 grant ANR-16-CE30-0019-02.

\bibliography{biblio}

\bibliographystyle{apsrev4-1}

\section*{Appendix}
\emph{Laser Doping.}
The doping is performed in an ultra-high vacuum chamber (pressure \SI{1e-9}{mbar}), where a puff of the precursor gas BCl$_3$ is injected onto the p-type (100) Si sample surface.
The substrate is then melted with a \SI{25}{ns} ultraviolet pulse generated by an excimer XeCl laser.
The boron diffuses into the liquid silicon and is incorporated substitutionally during the fast epitaxy (recrystallisation velocity about \SI{4}{m/s}), while the Cl creates volatile compounds with the atoms in the first atomic layers, resulting in a cleaner sample surface.
With our GILD setup, we have observed that the maximal active doping boron density saturates around \SI{4.7e21}{cm^{-3}}, the density being possibly limited by the pulse duration.
The layers studied in this paper are near such saturation.
Close to the saturation concentration the boron activation is impaired, and the ratio of interstitial boron to substitutional boron increased.
This effect is particularly evident in the thinner layers, where this saturation happens sooner, leading to more interstitial boron and more disorder.\\

\emph{Silicon resonator design.}
The practical implementation of the coupling to the input line differs in the two geometries: in the "transmission" coupled resonator, one can easily tune $Q_c$ from about \numrange{1e2}{1e6} by adjusting the input and output coupling capacitors.
For moderately low $Q_c$ such as the ones targeted here, we have implemented this coupling using interdigitated capacitors that have \SI{30}{\micro m} long fingers and \SI{4}{\micro m} spacing.
In the "shunt" coupling scheme, one generally tunes $Q_c$ with the length of the resonator conductor running along the transmission line (see Fig.~\ref{fig_GILD_setup_sample}b).
In that case, to obtain values of  $Q_c \sim 3000$, the coupling length may reach a few mm, comparable to the resonator length.
This results in less controlled coupling quality factors for the higher harmonics of the resonator.

Note that the "shunt" coupled has the advantage over the "transmission" coupled geometry to enable extracting both $Q_c$ and $Q_\mathrm{i}$ as independent parameters from a single measurement.
This is simply because the out-of-resonance signal provides a calibration of the measurement acquisition chain, while in the "transmission" scheme, one has to calibrate the setup in a separate measurement, or rely on microwave simulation for the knowledge of $Q_c$.
Note that in the "shunt" coupled geometry the fitted $Q_c$ were always within $10 \%$ of the simulated values, which gives us confidence in the reliability of the values of $Q_c$ used in the "transmission" geometry.\\

\emph{Extraction of the sheet kinetic inductance}
The analytical expression of the resonance frequency of the first harmonic reads $f_1 = 1/p l \sqrt{\mathcal{C}_g (\mathcal{L}_g+\mathcal{L}_k})$, where $l$ is the length of the resonator, $p = 2$ for a half-wave resonator, and $p = 4$ for a quarter-wave resonator.
The geometrical capacitance and inductance per unit length $\mathcal{C}_g$ ($\sim \SI{150}{pF/m}$) and $\mathcal{L}_g$ ($\sim \SI{500}{nH/m}$) are found from standard analytical expressions using the CPW dimensions and the dielectric constant of the Si substrate.
The sheet kinetic inductance $L_{k,\square}$ is calculated from the kinetic inductance per unit length as $\mathcal{L}_k = g_c L_{k,\square}^{SiB} + g_g L_{k,\square}^{Al}$, with $g_c$ and $g_g$  geometrical factors derived in e.g.~\cite{Gao2008} by integrating over current distributions in the CPW.
In our case, $\mathcal{L}_k  \approx g_c L_{k,\square}^{SiB}$, since the sheet kinetic inductance of the Al film is much smaller than the Si:B one.\\

\emph{Measurement setup and linear response.}
We perform the measurements in a cryogen-free dilution refrigerator with the temperature of the cold stage regulated between \SI{12}{mK} and \SI{2}{K}.
A \SIrange{1}{12}{GHz} signal generated by a vector network analyzer (VNA) is attenuated by \SIrange{90}{110}{dB} (depending on the experiment) by both room temperature and cold attenuators before reaching the sample.
The incoming power on the sample is known to \SI{1}{dB} from a separate calibration of the input line.
The signal leaving the sample is amplified by an amplifier chain with a total gain of approximately \SI{107}{dB} then fed to the input port of the VNA.
The first-stage HEMT amplifier anchored at \SI{4}{K} has a gain of \SI{32}{dB} and a noise temperature around \SI{9}{K} at \SI{3}{GHz}.
To protect the Si:B resonators from thermal radiation from the HEMT, either a \SIrange{3}{12}{GHz} isolator or an attenuator is anchored to the coldest stage between the sample and the HEMT.

In order to reach thermal equilibrium during the temperature ramp, at each temperature step the mixing chamber temperature is regulated from the indication of a thermometer anchored to the sample, and the sample is allowed to thermalize for 10 to 15 minutes before data acquisition is started.
At all temperatures, we used the highest admissible power to maintain linear response while improving the signal-to-noise ratio.\\

\emph{Losses from magnetic vortices} It is possible to estimate the complex resistivity associated with the vortex motion:
\begin{equation}
\rho_v = \rho_n \, \frac{B - B_{th}}{B_{c,2}} \; \frac{i f / f_d}{1+ i f / f_d}
\label{rv}
\end{equation}
where $\rho_n \sim 10^{-4}\,\Omega \,$cm is Si:B normal-state resistivity, $B_{c,2} \sim 0.1\,$T the upper critical field, and $f_d$ the characteristic depinning frequency \cite{Gittleman1968, Song2009}. $B_{th}$ is the threshold magnetic field for vortex generation, which scales with the width of the Si:B line as $B_{th} \sim \Phi_0/w^2$, giving $B_{th}\sim 0.8\,$G for $w=5\,\mu$m (sample C).
This threshold is about three orders of magnitude higher than the microwave-induced magnetic field; however, vortices can be present due to imperfect magnetic screening of the sample.
For frequencies smaller than the depinning frequency $f_d$, the real dissipative, part of the complex resistivity goes as $\rho_v \sim f^2$, as qualitatively observed in Fig.~\ref{fig_summary_linear}g.
Moreover, a quantitative agreement can be obtained with eq.\ref{rv}. The depinning frequency calculated from the experimental DC parameters is $f_d = \rho_n j_c / 2 \pi \xi B_{c,2} = 19$ GHz with $j_c \sim 6\times 10^8$ A/m$^2$ the measured critical current density. The threshold magnetic field was taken as $B_{th}= \Phi_0/w^2$. To interpret the measurements within this model we need to assume the presence of a stray $1-2\,$G magnetic field.\\

\emph{Non-linear response: resolution of the non-linear hamiltonian}
The steady-state equation for the electromagnetic energy inside the resonator is a third order polynomial which depends on incident power, frequency, and losses. This polynomial accepts between one and three real roots which are used to compute the forward scattering parameter $S_{21}(f)$.
Note that when there exists three real solutions for the energy, one is a low amplitude, one is a high amplitude, and the third is a metastable state not accessible experimentally.
The two accessible states describe the hysteretic regime, and are accessed depending on the history (eg frequency sweep up or down).\\
The presence of multiple harmonics in our resonators allows us to follow the evolution of $K$ and $\gamma$ with the frequency in the same device. 

\emph{Quasiparticle heating model}
The power dissipated by the QPs,
$P_{QP} = \overline{n} \hbar \omega_r \chi \kappa_{\mathrm{i}}$, is
proportional to the energy stored in the resonator $ \overline{n} \hbar \omega_r$ and to $ \chi \kappa_{\mathrm{i}} $ the energy loss rate due to quasiparticles, where $0 \le \chi \le 1$.
Thus, $P_{QP} = 2 P_{in}\,\chi\,\kappa_c\, \kappa_{\mathrm{i}}/\kappa_t^2\,(1 + (2 \delta \omega / \kappa_t)^2 ) $ is maximal at resonance ($\delta \omega=0$ ), and for critical coupling ($\kappa_c = \kappa_{\mathrm{i}}$).
Fixing $\chi$, one finds the equilibrium temperature by equating $P_{QP}$ with the quasiparticle-phonon power flow, given by the recombination term, proportional to the material-dependent $\Sigma_s$, and the scattering term, proportional to $\Sigma_n$ \cite{Guruswamy_th_2018}. From the equilibrium temperature, the resonance frequency and internal quality factor are found as a function of microwave power, which enables computing the apparent "Kerr constant" $\partial{f_r}/{\overline{n}}$.
\end{document}